\documentclass[10pt,aps,fleqn,superscriptaddress,notitlepage,nofootinbib,preprintnumbers,nobalancelastpage]{revtex4-1}
\pdfoutput=1
\usepackage{amsmath,amssymb,graphicx,xspace,subfigure,tikz}
\usetikzlibrary{calc}
\usepackage{todonotes}
\definecolor{lellow}{RGB}{255,200,0}

\usepackage{hyperref}
\hypersetup{
  pdfauthor={Stefan Hoeche, Daniel Reichelt},
  pdftitle={Numerical resummation at subleading color in the strongly ordered soft gluon limit},
  pdfkeywords={QCD, Resummation, Parton Shower}
}
\begin{document}
\preprint{FERMILAB-PUB-20-028-T, MCNET-20-06}
\title{Numerical resummation at subleading color in the strongly ordered soft gluon limit}
\author{Stefan~H{\"o}che}
\affiliation{Fermi National Accelerator Laboratory,
  Batavia, IL, 60510, USA}
\author{Daniel~Reichelt}
\affiliation{Institut f{\"u}r Theoretische Physik,
  Georg-August-Universit{\"a}t G{\"o}ttingen, D-37077 G{\"o}ttingen, Germany}
\begin{abstract}
  We present a Monte Carlo approach to soft-gluon resummation at subleading
  color which can be used to improve existing parton shower algorithms.
  At the single-emission level, soft-collinear enhancements of the splitting
  functions are explicitly linked to quadratic Casimir operators, while
  wide angle single-soft enhancements are connected to nontrivial color
  correlators. We focus on a numerically stable implementation of color
  matrix element corrections to all orders and approximate the virtual corrections
  by requiring unitarity at the single-emission level. We provide a proof-of-concept
  implementation to compute nonglobal event shapes at lepton colliders.
\end{abstract}
\maketitle

\section{Introduction}
Soft-gluon resummation is one of the most important tools in perturbative QCD,
as it allows one to systematically and fairly straightforwardly compute radiative
corrections to all orders for a large class of observables~\cite{Bassetto:1984ik}.
The effect of gluon radiation is typically computed for single or multiple
emissions, and recoil effects are approximated at the same level.\footnote{
  This is known to cause potentially sizable differences between the results
  from parton showers and analytic resummation~\cite{Hoeche:2017jsi}.}
If the observable is simple, all-order corrections can be obtained by exponentiating
these results, and the remaining obstacle is posed by color coherence,
which may lead to a soft anomalous dimension in matrix form.
A general framework for resumming event shapes based on this concept
was developed at next-to-leading logarithmic (NLL) accuracy~\cite{
  Banfi:2001bz,Banfi:2003je,Banfi:2004nk,Banfi:2004yd} and at
next-to-next-to-leading logarithmic accuracy~\cite{Banfi:2014sua,Banfi:2016zlc}.

nonglobal observables require a more sophisticated treatment,
which was first discussed in the context of $e^+e^-$ and deep inelastic
scattering (DIS) event shape
resummation~\cite{Dasgupta:2001sh,Dasgupta:2002dc} using a Monte Carlo (MC)
approach at leading color accuracy~\cite{Dasgupta:2001sh}.
A corresponding evolution equation was derived~\cite{Banfi:2002hw},
which enabled the inclusion of subleading color effects~\cite{Weigert:2003mm,
  Caron-Huot:2015bja}. Numerical results have subsequently been computed
for example for interjet energy flows~\cite{Hatta:2013iba} and for the
hemisphere mass distribution in $e^+e^-\to$hadrons~\cite{Hagiwara:2015bia}.
nonglobal observables have also been investigated using methods of
effective field theories~\cite{Larkoski:2015zka,Becher:2015hka,Becher:2016mmh}. 

nonglobal logarithms are particularly important in the context of
interjet radiation and in the presence of a jet veto
\cite{Forshaw:2006fk, Forshaw:2008cq}. In the context of Large Hadron Collider
phenomenology they have therefore received considerable attention
\cite{Forshaw:2009fz,Delgado:2011tp,Banfi:2012yh,Banfi:2012jm,Banfi:2015pju}.
Several approaches have been suggested for their numerical resummation, ranging
from subleading color parton showers~\cite{Gustafson:1992uh,Platzer:2012np,
  Platzer:2018pmd,Nagy:2012bt,Nagy:2019pjp,Hamilton:2020rcu}
to evolution at the amplitude level~\cite{Martinez:2018ffw,Forshaw:2019ver,
  DeAngelis:2020rvq,Holguin:2020joq,Platzer:2020lbr}.
While color-corrected parton showers can exhibit good numerical convergence,
the evaluation of the color matrix elements becomes prohibitively expensive
at high parton multiplicity, and therefore the approach cannot be used beyond
a very limited number of emissions. Amplitude-level evolution on the other hand
will typically suffer from a slow rate of convergence in the Monte Carlo simulation.
In order to address the problem of slow convergence, we propose a novel algorithm.
Using color conservation, the squared soft-gluon current
is rearranged into a soft-collinear contribution proportional to the quadratic
Casimir operator, and a collinearly suppressed correction term proportional
to the color correlators. Based on the independence of color and kinematics
operators, the color matrix elements are integrated with Monte Carlo methods
at each step of the evolution. This allows one to reach good precision on the
color coefficients, while limiting the run-time of computer simulations at high
multiplicity. Appendix~\ref{sec:caesar} shows how color coherence emerges in
our approach.

This manuscript is organized as follows: Section~\ref{sec:resummation} discusses
the resummation formalism, and Sec.~\ref{sec:ps_map} introduces the phase-space
mapping needed to implement it away from the exact soft limit.
Section~\ref{sec:color_sampling} presents our Monte Carlo technique to compute
the color matrix elements. The difference between subleading color and improved leading
color evolution used in standard parton showers is analyzed in Sec.~\ref{sec:numerics}
by studying the light jet mass and narrow jet broadening distributions in
$e^+e^-\to$hadrons. Section~\ref{sec:conclusions} contains an outlook.

\section{Resummation formalism}
\label{sec:resummation}
Soft-gluon resummation is typically performed for a given, fixed number
of hard partons, generated at scales that are widely separated from the scale
of additional soft radiation. These partons are assumed to be unchanged after
the emission of a soft gluon, leading to the notion of Wilson lines and
eventually the exponentiation of the soft anomalous dimension matrix. 
We will adopt a different approach, based on the physical picture
in the strongly ordered soft limit. By the very definition of strong
ordering, each radiated soft gluon must be treated as a new Wilson line
for subsequent gluon emissions.

We denote the Born matrix element for $n$ partons by $|M_n\rangle$
and the color insertion operator for parton $i$ as ${\bf T}_i$~\cite{Bassetto:1984ik}.
The approximate $n+1$-parton squared matrix element in the soft limit then reads
\begin{equation}\label{eq:soft_me2}
  \langle m_{n+1}|m_{n+1}\rangle = \langle M_n|{\bf\Gamma}_n({\bf 1})|M_n\rangle\;,
\end{equation}
where we have defined the squared $n$-parton soft current
\begin{equation}\label{eq:soft_insertion2}
  {\bf\Gamma}_n({\bf \Gamma})=-\sum_{i=1}^{n}\sum_{\substack{j=1\\j\neq i}}^n
  {\bf T}_i\,{\bf\Gamma}\,{\bf T}_j\,w_{ij}\;,
  \qquad\text{with}\qquad
  w_{ij}=\frac{s_{ij}}{s_{iq}s_{jq}}\;.
\end{equation}
The invariants $s_{ij}$ are defined in terms of the parton momenta, $p_i$,
as $s_{ij}=2p_ip_j$. Since $|m_{n+1}\rangle$ is an $n+1$-parton matrix element,
$\Gamma$ is defined in a higher dimensional color space than $\Gamma_n$.
Equation~\eqref{eq:soft_me2} generalizes to $k+1$ emissions as
\begin{equation}\label{eq:soft_me2_k}
  \langle m_{n+k+1}|m_{n+k+1}\rangle =
  \langle m_{n+k}|{\bf\Gamma}_{n+k}({\bf 1})|m_{n+k}\rangle =
  \langle M_n|{\bf\Gamma}_{n}({\bf \Gamma}_{n+1}(...\,
    {\bf\Gamma}_{n+k-1}({\bf\Gamma}_{n+k}({\bf 1}))...))|M_n\rangle\;.
\end{equation}
Note that the complexity of $\Gamma_n$ increases rapidly with the number
of partons, such that the evaluation of color factors encoded in
Eq.~\eqref{eq:soft_me2_k} becomes increasingly cumbersome.
The evolution of the parton ensemble is governed by the differential
branching probability
\begin{equation}\label{eq:branching_probability}
  \frac{{\rm d}\sigma_{n+k+1}}{\sigma_{n+k}}=
  {\rm d}\Phi_{+1}\,8\pi\alpha_s\,
  \frac{\langle m_{n+k}|{\bf\Gamma}_{n+k}({\bf 1})|m_{n+k}\rangle}{
    \langle m_{n+k}|m_{n+k}\rangle}\;,
\end{equation}
where ${\rm d}\Phi_{+1}={\rm d}^4q\,\delta(q^2)/(2\pi)^3$ is the four-dimensional
differential phase-space element for the emission of the gluon with momentum $q$.
We parametrize this phase space as
\begin{equation}\label{eq:ps_parametrization}
  {\rm d}\Phi_{+1}=\frac{1}{16\pi^2}\,
  {\rm d}\kappa^2_{ij}\,{\rm d}\tilde{z}_i\,\frac{{\rm d}\phi_{ij}}{2\pi}\,
    J(\kappa^2_{ij},\tilde{z}_i,\phi_{ij,m})\;,
\end{equation}
where $\kappa^2_{ij}$ is the evolution variable of the parton shower,
$\tilde{z}_i$ is the splitting variable, $\phi_{ij}$ is an azimuthal angle,
and $J(\kappa^2_{ij},\tilde{z}_i,\phi_{ij,m})$ is a Jacobian factor.
For details, see Sec.~\ref{sec:ps_map}.

Equation~\eqref{eq:branching_probability} describes resolved real-emission corrections.
A standard choice for parton-shower algorithms is to define a no-branching probability,
$\Pi(\kappa^2)$, such that virtual and unresolved real-emission corrections are defined
in terms of the resolved real-emission corrections by means of unitarity.
This approach should be improved by accounting for the different color structure
in virtual corrections~\cite{Martinez:2018ffw,Forshaw:2019ver,Nagy:2019pjp},
as well as for Coulomb phases~\cite{Martinez:2018ffw,Forshaw:2019ver,Nagy:2019rwb}.
It was pointed out in~\cite{Holguin:2020xx} that one can therefore not yet claim a fully
color correct evolution. We will postpone these problems to a future publication.
Instead we focus our attention on a suitable rearrangement of color and kinematics factors
in the real components, in order to improve the numerical convergence of the simulation.
While this is not sufficient for arbitrarily complicated observables, it constitutes an
important step toward a complete full-color resummation algorithm.

We define the no-branching probability such as to restore unitarity
\begin{equation}
  \int_{t}^{Q^2}{\rm d}\kappa_{ij}^2\,\frac{1}{\sigma_{n+k}}
  \int\frac{{\rm d}\sigma_{n+k+1}}{{\rm d}\kappa_{ij}^2}\,
  \Pi(\kappa_{ij}^2,Q^2,{\bf 1})=1-\Pi(t,Q^2,{\bf 1})\;.
\end{equation}
This equation has the solution
\begin{equation}\label{eq:full_sudakov}
  \Pi(t,Q^2,{\bf \Gamma})
  =\prod_{i=1}^{n}\prod_{\substack{j=1\\j\neq i}}^n
  \Pi_{ij}(t,Q^2,{\bf \Gamma})\;.
\end{equation}
where
\begin{equation}\label{eq:single_sudakov}
  \begin{split}
  \Pi_{ij}(t,Q^2,{\bf \Gamma})
  =&\;\exp\left\{-\int_{t}^{Q^2}\frac{{\rm d}\kappa_{ij}^2}{\kappa_{ij}^2}
  \int{\rm d}\tilde{z}_i\int\frac{{\rm d}\phi_{ij}}{2\pi}\,
  J(\kappa^2_{ij},\tilde{z}_i,\phi_{ij})\,\frac{\alpha_s}{2\pi}\,
  \frac{\langle m_{n+k}|{\bf T}_i\,{\bf\Gamma}\,{\bf T}_j|m_{n+k}\rangle}{
    \langle m_{n+k}|m_{n+k}\rangle}\,\right\}\;.
  \end{split}
\end{equation}
The squared $n$-parton soft current, Eq.~\eqref{eq:soft_insertion2}
has a form which is not particularly suitable for implementation
in numerical simulations. We use the partial fractioning approach
of~\cite{Catani:1996vz} to rearrange it as
\begin{equation}\label{eq:soft_ps}
  {\bf\Gamma}_n({\bf\Gamma})=
  -\frac{1}{2}\sum_{i=1}^{n}\sum_{\substack{j=1\\j\neq i}}^n{\bf T}_i\,{\bf\Gamma}\,{\bf T}_j(P^i_j+P^j_i)
  =-\frac{1}{2}\sum_{i=1}^{n}\sum_{\substack{j=1\\j\neq i}}^n
  \Big({\bf T}_i\,{\bf\Gamma}\,{\bf T}_j+{\bf T}_j\,{\bf\Gamma}\,{\bf T}_i\Big)P^i_j\;,
\end{equation}
where we have defined the splitting operator
\begin{equation}\label{eq:def_pij}
  P^i_j=\frac{1}{s_{iq}}\frac{2\,s_{ij}}{s_{iq}+s_{jq}}\;.
\end{equation}
Note that in general ${\bf T}_i{\bf\Gamma}{\bf T}_j$ will not
equal ${\bf T}_j{\bf\Gamma}{\bf T}_i$, hence we cannot combine
the two terms on the right-hand side of Eq.~\eqref{eq:soft_ps}.
We use color conservation to rewrite them as
\begin{equation}
  \begin{split}
    {\bf\Gamma}_n({\bf\Gamma})=&\;\frac{1}{n-1}\sum_{i=1}^n
    \sum_{\substack{j=1\\j\neq i}}^n
    \bigg({\bf T}_i\,{\bf\Gamma}\,{\bf T}_iP^i_j
    +\frac{1}{2}\sum_{\substack{k=1\\k\neq i,j}}^n
    \Big({\bf T}_i\,{\bf\Gamma}\,{\bf T}_k
    +{\bf T}_k\,{\bf\Gamma}\,{\bf T}_i\Big)P^i_j
    -\frac{n-2}{2}\,\Big({\bf T}_i\,{\bf\Gamma}\,{\bf T}_j
    +{\bf T}_j\,{\bf\Gamma}\,{\bf T}_i\Big)P^i_j\,\bigg)\;.
  \end{split}
\end{equation}
Combining the second and the last terms in parentheses, we obtain
\footnote{In the special case of ${\bf\Gamma}={\bf 1}$,
  i.e.\ at fixed order, we can simplify Eq.~\eqref{eq:coll_sub_soft}
  using the identity ${\bf T}_i^2=C_i$. This relation was used
  to reformulate the two-loop soft function in~\cite{Dulat:2018vuy}.}
\begin{equation}\label{eq:coll_sub_soft}
  \begin{split}
    {\bf\Gamma}_n({\bf\Gamma})=&\;\frac{1}{n-1}\sum_{i=1}^n
    \sum_{\substack{j=1\\j\neq i}}^n
    \bigg({\bf T}_i\,{\bf\Gamma}\,{\bf T}_iP^i_j
    +\frac{1}{2}\sum_{\substack{k=1\\k\neq i,j}}^n
    \Big({\bf T}_i\,{\bf\Gamma}\,{\bf T}_k+
        {\bf T}_k\,{\bf\Gamma}\,{\bf T}_i\Big)\tilde{P}^i_{jk}\bigg)\;,
  \end{split}
\end{equation}
where we have defined the splitting operator~\cite{Dulat:2018vuy}
\begin{equation}\label{eq:def_ptilde}
  \tilde{P}^i_{jk}=P^i_j-P^i_k\;.
\end{equation}
Note that $\tilde{P}$ tends to zero in the $iq$-collinear limit.
Equation~\eqref{eq:coll_sub_soft} should therefore be viewed
as a rearrangement of Eq.~\eqref{eq:soft_ps}, where the collinearly
enhanced terms are made explicit, and the remainder is singly soft
enhanced only. While additional rearrangements would allow one to achieve
a further kinematical suppression by combining multiple operators as
$\tilde{P}^i_{jk}+\tilde{P}^j_{il}+\tilde{P}^k_{li}+\tilde{P}^l_{kj}$,
such rearrangements will produce additional terms proportional to
${\bf T}_i\,{\bf\Gamma}\,{\bf T}_i$. We find
Eq.~\eqref{eq:coll_sub_soft} to be the most suitable form
for a Monte Carlo implementation. Examples for its relation to
analytically known soft insertion operators are given in
Appendix~\ref{sec:examples}.

\section{Kinematics mapping}
\label{sec:ps_map}
\begin{figure}[t]
  \centerline{\includegraphics[width=\textwidth]{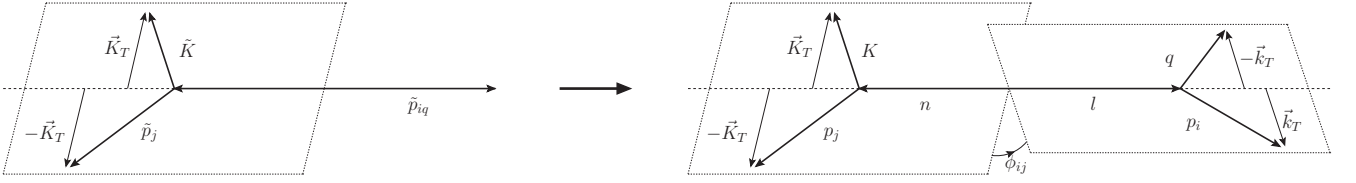}}
  \caption{Sketch of the kinematics mapping described in Sec.~\ref{sec:ps_map}.
    The emitting parton is $i$, and the reference momentum for definition
    of the azimuthal angle is $j$. The emitted gluon carries momentum $q$.
    The forward and backward light-cone momenta are given by $l$ and $n$.}
  \label{fig:ps_mapping}
\end{figure}
In order to implement Eq.~\eqref{eq:coll_sub_soft}
in a numerical simulation, the operator $\tilde{P}^i_{jk}$ must be well defined.
When evaluating the difference between $P^i_k$ and $P^i_j$, we assume that either the
underlying Born configurations are identical in both terms, or that their
difference gives rise to subleading power corrections. Since the latter may
be difficult to prove in the general case, we use a kinematics mapping,
which ensures that the underlying Born state is the same for identical
$i$ and $q$. Such a mapping is defined, for example in~\cite{
  Nagy:2007ty,Nagy:2008ns,Nagy:2008eq,Nagy:2014mqa,
  Chung:2010fx,Chung:2012rq}, and is schematically depicted
in Fig.~\ref{fig:ps_mapping}. Here and in the following,
a tilde denotes momenta before the emission of the soft gluon.

We define the variables in Eq.~\eqref{eq:ps_parametrization} as follows
\begin{equation}\label{eq:def_ps_variables}
  \begin{split}
  \kappa_{ij}^2=&\;\frac{s_{iq}s_{jq}}{s_{ij}}\;,\qquad
  \tilde{z}_i=\frac{s_{ij}}{s_{ij}+s_{jq}}\;,\\
  2\cos\phi_{ij}=&\;\sqrt{\frac{s_{ni}s_{nq}}{s_{iq}}}
  \sqrt{\frac{s_{nj}s_{nm}}{s_{jm}}}\left(
  \frac{s_{im}}{s_{nj}s_{nq}}+\frac{s_{jq}}{s_{ni}s_{nm}}
  -\frac{s_{ij}}{s_{nm}s_{nq}}-\frac{s_{mq}}{s_{ni}s_{nj}}\right)\;,
  \end{split}
\end{equation}
where the lightlike reference vector $n^\mu$ is defined in Eq.~\eqref{eq:sudakov_parametrization}.
The lightlike auxiliary vector $m^\mu$ is given by $m^\mu=K-K^2/(2\,Kn)\,n$ and defines the
reference axis for $\phi_{ij}$ in the transverse plane.
Note that the inverse mapping leads to the same $n+k$-parton momentum configuration for
any choice of $p_j^\mu$ and $m^\mu$, as long as $p_i^\mu$ and $q$ are identical.
This is an important feature needed for the rearrangement of color insertion operators
in Sec.~\ref{sec:resummation}.

The longitudinal recoil generated in the emission of the
soft gluon $q$ is absorbed by all partons except the emitter.
Due to our choice of evolution and splitting variables,
the mapping depends nontrivially on the azimuthal angle.
In order to construct the momenta, we first define the
lightlike vectors
\begin{equation}\label{eq:sudakov_parametrization}
  l^\mu=\frac{1+s_{jK}/\tilde{\gamma}}{1+s_{jK}/\gamma}\,\tilde{p}_{iq}^\mu\;,\qquad
  n^\mu=\frac{1}{1+s_{iq}/\gamma}\bigg(\tilde{K}^\mu+\tilde{p}_j^\mu-\frac{(\tilde{K}+\tilde{p}_j)^2}{\gamma}\,\tilde{p}_{iq}^\mu\bigg)\;,
\end{equation}
where
\begin{equation}
  \tilde{\gamma}=2(\tilde{K}+\tilde{p}_j)\cdot\tilde{p}_{iq}=Q^2-(\tilde{K}+\tilde{p}_j)^2\;.
\end{equation}
The rescaled invariant $\gamma$ is given by
\begin{equation}
  \gamma=\frac{1}{2}\Big(\big(Q^2-(\tilde{K}+\tilde{p}_j)^2-s_{iq}\big)
  +\sqrt{(Q^2-(\tilde{K}+\tilde{p}_j)^2-s_{iq})^2-4(\tilde{K}+\tilde{p}_j)^2s_{iq}}\;\Big)\;.
\end{equation}
We can now parametrize the momenta as
\begin{equation}
  \begin{split}
    p_i^\mu=&\;z\,l^\mu+\frac{\vec{k}_T^2}{z\gamma}\,n^\mu+k_T^\mu\;,
    \qquad& q^\mu=&\;(1-z)\,l^\mu+\frac{\vec{k}_T^2}{(1-z)\gamma}\,n^\mu-k_T^\mu\;,
  \end{split}
\end{equation}
and
\begin{equation}
  \begin{split}
    p_j^\mu=&\;x\,n^\mu+\frac{\vec{K}_T^2}{x\gamma}\,l^\mu-K_T^\mu\;,
    \qquad& K^\mu=&\;(1-x)\,n^\mu+\frac{\tilde{K}^2+\vec{K}_T^2}{(1-x)\gamma}\,l^\mu+K_T^\mu\;.
  \end{split}
\end{equation}
Note that $x$ and $K_T$ are invariant under the mapping, as the
momenta $p_j$ and $K$ are completely determined by a boost of
$\tilde{p}_j$ and $\tilde{K}$ along the direction of $n$.
The variables $x$ and $\vec{K}_T^2$ can therefore be computed
using the Born kinematics. Solving Eqs.~\eqref{eq:def_ps_variables}
for $z$ and $\vec{k}_T^2$ then yields
\begin{equation}
  \begin{split}
    z=&\;\frac{\tilde{z}_iC+(1-\tilde{z}_i)/C+2\cos^2\phi_{ij}-1}{
      (C-1)^2/C+4\cos^2\phi_{ij}}+{\rm sgn}(\cos\phi_{ij})\,
    \frac{\sqrt{(C+1)^2/C\,\tilde{z}_i(1-\tilde{z}_i)-\sin^2\phi_{ij}}}{
      (C-1)^2/C+4\cos^2\phi_{ij}}\;,\\
    \vec{k}_T^2=&\;\kappa_{ij}^2\frac{\tilde{z}_i}{1-\tilde{z}_i}\,z(1-z)\;,
    \qquad\text{where}\qquad
    C=\frac{(1-\tilde{z}_i)x^2}{\tilde{z}_i\vec{K}_T^2\kappa_{ij}^2}\;.
  \end{split}
\end{equation}
The phase-space boundaries are given by
\begin{equation}
  s_{iq}\leq Q^2-\tilde{K}^2\;
  \qquad\text{and}\qquad
  z < \Big[\,1+{\rm sgn}(\cos\phi_{ij})\sqrt{\vec{K}_T^2\kappa_{ij}^2}/(x\gamma)\,\Big]^{-1}\;.
\end{equation}
The Jacobian introduced in Eq.~\eqref{eq:ps_parametrization} is given by
\begin{equation}
  J(\kappa_{ij}^2,\tilde{z}_i,\phi_{ij})=
  \frac{\sqrt{(Q^2-\tilde{K}^2-s_{iq})^2-4\tilde{K}^2s_{iq}}}{Q^2-\tilde{K}^2}\,
  \frac{2z(1-z)\big(z(1-z)\,x^2\gamma^2+\vec{K}_T^2\vec{k}_T^2\big)}{
    z(1-z)\,x\gamma\,\big(zs_{jq}+(1-z)s_{ij}\big)-\vec{K}_T^2\vec{k}_T^2}\,
  \frac{\kappa_{ij}^2}{\vec{k}_T^2}\;.
\end{equation}
The fact that the inverse of this mapping yields the same underlying Born
kinematics for all configurations where the emitting particle is parton $i$
allows one to rearrange the soft anomalous dimension matrix ${\bf\Gamma}$ into
Eq.~\eqref{eq:coll_sub_soft} without the need for an additional
reweighting away from the exact soft limit.

\section{Color algebra}
\label{sec:color_sampling}
\begin{figure}[t]
  \centerline{\includegraphics[width=\textwidth]{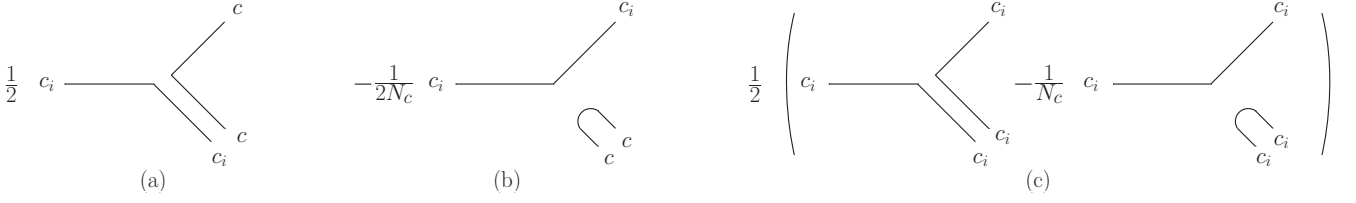}}
  \caption{Sketch of color flow in quark to quark gluon splittings.
    The original quark color state is $(c_i,0)$. In case (a), the newly
    generated color $c$ differs from $c_i$, and the gluon is assigned
    the color state $(c_i,c)$. In case (b), the newly generated color
    $c$ differs from $c_i$ and the gluon is assigned the color state $(c,c)$.
    In case (c), the newly generated color is the same as the original color,
    $c_i$. The prefactors indicate the weight of each diagram,
    which originates in the Fierz identity
    $T^a_{ik}T^a_{mn}=1/2(\delta_{in}\delta_{mk}-\delta_{ik}\delta_{mn}/N_c)$
    and appears squared in the sampling algorithm. See the main text for details.}
  \label{fig:color_sampling}
\end{figure}
\begin{table}[t]
  \begin{tabular}{l|c|c}\cline{1-3}
    \vphantom{$\displaystyle\int_A^B$}Coefficient & Analytic value / $N_c$ & MC result / $N_c$\\\cline{1-3}
    \begin{tikzpicture}[baseline=-\the\dimexpr\fontdimen22\textfont2\relax]
      \def\Radius{0.5}
      \draw (0,0) circle (\Radius);
      \draw[color=blue] (60:0) -- (60:\Radius);
      \draw[color=red] (300:0) -- (300:\Radius);
      \draw[color=green] (180:0) -- (180:\Radius);
      \useasboundingbox ([shift={(1mm,1mm)}]current bounding box.north east)
            rectangle ([shift={(-1mm,-1mm)}]current bounding box.south west);
    \end{tikzpicture}= $F^c_{ab}{\rm Tr}\left[T^aT^bT^c\right]$
    &$\displaystyle C_F\frac{C_A}{2}$&1.9998(2)\\
    \begin{tikzpicture}[baseline=-\the\dimexpr\fontdimen22\textfont2\relax]
      \def\Radius{0.5}
      \draw (0,0) circle (\Radius);
      \draw[color=blue] (0:-\Radius/3) -- (135:\Radius);
      \draw[color=magenta] (0:-\Radius/3) -- (225:\Radius);
      \draw[color=cyan] (0:-\Radius/3) -- (0:\Radius/3);
      \draw[color=green] (0:\Radius/3) -- (45:\Radius);
      \draw[color=red] (0:\Radius/3) -- (-45:\Radius);
      \useasboundingbox ([shift={(1mm,1mm)}]current bounding box.north east)
            rectangle ([shift={(-1mm,-1mm)}]current bounding box.south west);
    \end{tikzpicture}= $F^d_{ae}F^c_{eb}{\rm Tr}\left[T^aT^bT^cT^d\right]$
    &$\displaystyle C_F\left(\frac{C_A}{2}\right)^2$&2.9995(4)\\
    \begin{tikzpicture}[baseline=-\the\dimexpr\fontdimen22\textfont2\relax]
      \def\Radius{0.5}
      \def\Rot{90}
      \def\conna{-108}
      \def\scalea{2.2}
      \def\scaleb{3.5}
      \draw (0,0) circle (\Radius);
      \draw[color=magenta] (\Rot-\conna:\Radius/\scalea) -- (\Rot+144:\Radius);
      \draw[color=cyan] (\Rot+0:\Radius/\scaleb) -- (\Rot+\conna:\Radius/\scalea);
      \draw[color={rgb,255:red,20;green,170;blue,170}] (\Rot-\conna:\Radius/\scalea) -- (\Rot+0:\Radius/\scaleb);
      \draw[color=blue] (\Rot-\conna:\Radius/\scalea) -- (\Rot+72:\Radius);
      \draw[color=green] (\Rot+0:\Radius/\scaleb) -- (\Rot+0:\Radius);
      \draw[color=lellow] (\Rot+\conna:\Radius/\scalea) -- (\Rot-72:\Radius);
      \draw[color=red] (\Rot+\conna:\Radius/\scalea) -- (\Rot-144:\Radius);
      \useasboundingbox ([shift={(1mm,1mm)}]current bounding box.north east)
            rectangle ([shift={(-1mm,-1mm)}]current bounding box.south west);
    \end{tikzpicture}= $F^e_{af}F^d_{fg}F^c_{gb}{\rm Tr}\left[T^aT^bT^cT^dT^e\right]$
    &$\displaystyle C_F\left(\frac{C_A}{2}\right)^3$&4.4996(8)\\
    \begin{tikzpicture}[baseline=-\the\dimexpr\fontdimen22\textfont2\relax]
      \def\Radius{0.5}
      \def\Rot{0}
      \def\CSep{2.25}
      \draw (0,0) circle (\Radius);
      \draw[color=lellow] (\Rot:\Radius/\CSep) -- (\Rot+90:\Radius/\CSep);
      \draw[color=blue] (\Rot+180:\Radius/\CSep) -- (\Rot+90:\Radius/\CSep);
      \draw[color={rgb,255:red,141;green,0;blue,231}] (\Rot+90:\Radius/\CSep) -- (\Rot+90:\Radius);
      \draw[color=cyan] (\Rot+180:\Radius/\CSep) -- (\Rot-90:\Radius/\CSep);
      \draw[color=green] (\Rot-90:\Radius/\CSep) -- (\Rot:\Radius/\CSep);
      \draw[color={rgb,255:red,20;green,170;blue,170}] (\Rot:\Radius/\CSep) -- (\Rot:\Radius);
      \draw[color=magenta] (\Rot+180:\Radius/\CSep) -- (\Rot+180:\Radius);
      \draw[color=red] (\Rot-90:\Radius/\CSep) -- (\Rot-90:\Radius);
      \useasboundingbox ([shift={(1mm,1mm)}]current bounding box.north east)
            rectangle ([shift={(-1mm,-1mm)}]current bounding box.south west);
    \end{tikzpicture}= $F^e_{ag}F^f_{gb}F^f_{ch}F^e_{hd}{\rm Tr}\left[T^aT^bT^cT^d\right]$
    &$\displaystyle C_F\left(\frac{C_A}{2}\right)^3\left(1+\frac{2}{N_c^2}\right)$&5.499(1)\\
    \cline{1-3}
    \begin{tikzpicture}[baseline=-\the\dimexpr\fontdimen22\textfont2\relax]
      \def\Radius{0.5}
      \draw (0,0) circle (\Radius);
      \draw[color=blue] (30:-\Radius) -- (90:\Radius);
      \draw[color=green] (30:0) -- (30:\Radius);
      \draw[color=red] (270:0) -- (270:\Radius);
      \draw[color=cyan] (150:0) -- (150:\Radius);
      \useasboundingbox ([shift={(1mm,1mm)}]current bounding box.north east)
            rectangle ([shift={(-1mm,-1mm)}]current bounding box.south west);
    \end{tikzpicture}= $F^d_{bc}{\rm Tr}\left[T^aT^bT^aT^cT^d\right]$
    &$\displaystyle -C_F\frac{C_A}{2}\left(\frac{C_A}{2}-C_F\right)$&-0.3332(3)\\
    \begin{tikzpicture}[baseline=-\the\dimexpr\fontdimen22\textfont2\relax]
      \def\Radius{0.5}
      \def\Rot{0}
      \draw (0,0) circle (\Radius);
      \draw[color=blue] (\Rot+180:\Radius) -- (\Rot+270:\Radius);
      \draw[color=green] (\Rot:-\Radius/3) -- (\Rot+135:\Radius);
      \draw[color=magenta] (\Rot:-\Radius/3) -- (\Rot+225:\Radius);
      \draw[color=cyan] (\Rot:-\Radius/3) -- (\Rot:\Radius/3);
      \draw[color=lellow] (\Rot:\Radius/3) -- (\Rot+45:\Radius);
      \draw[color=red] (\Rot:\Radius/3) -- (\Rot-45:\Radius);
      \useasboundingbox ([shift={(1mm,1mm)}]current bounding box.north east)
            rectangle ([shift={(-1mm,-1mm)}]current bounding box.south west);
    \end{tikzpicture}= $F^e_{bf}F^d_{fc}{\rm Tr}\left[T^aT^bT^aT^cT^dT^e\right]$
    &$\displaystyle -C_F\left(\frac{C_A}{2}\right)^2\left(\frac{C_A}{2}-C_F\right)$&-0.5001(5)\\
    \begin{tikzpicture}[baseline=-\the\dimexpr\fontdimen22\textfont2\relax]
      \def\Radius{0.5}
      \def\Rot{0}
      \draw (0,0) circle (\Radius);
      \draw[color=lellow] (\Rot+30:\Radius) -- (\Rot+150:\Radius);
      \draw[color=blue] (\Rot+225:\Radius/3) -- (\Rot+104:\Radius);
      \draw[color=magenta] (\Rot+225:\Radius/3) -- (\Rot+225:\Radius);
      \draw[color=cyan] (\Rot+225:\Radius/3) -- (\Rot-45:\Radius/3);
      \draw[color=green] (\Rot-45:\Radius/3) -- (\Rot+76:\Radius);
      \draw[color=red] (\Rot-45:\Radius/3) -- (\Rot-45:\Radius);
      \useasboundingbox ([shift={(1mm,1mm)}]current bounding box.north east)
            rectangle ([shift={(-1mm,-1mm)}]current bounding box.south west);
    \end{tikzpicture}= $F^e_{bf}F^d_{fc}{\rm Tr}\left[T^aT^bT^cT^aT^dT^e\right]$
    &$\displaystyle -C_F\left(\frac{C_A}{2}\right)^2\left(\frac{C_A}{2}-C_F-\frac{C_A}{N_c^2}\right)$&0.5007(4)\\
    \cline{1-3}
    \begin{tikzpicture}[baseline=-\the\dimexpr\fontdimen22\textfont2\relax]
      \def\Radius{0.5}
      \draw (0,0) circle (\Radius);
      \draw[color=blue] (45:-\Radius) -- (90:\Radius);
      \draw[color=green] (0:-\Radius) -- (45:\Radius);
      \draw[color=cyan] (-45:-\Radius) -- (-45:0);
      \draw[color=lellow] (-45:0) -- (-15:\Radius);
      \draw[color=red] (-45:0) -- (-75:\Radius);
      \useasboundingbox ([shift={(1mm,1mm)}]current bounding box.north east)
            rectangle ([shift={(-1mm,-1mm)}]current bounding box.south west);
    \end{tikzpicture}= $F^e_{cd}{\rm Tr}\left[T^aT^bT^cT^aT^bT^dT^e\right]$
    &$\displaystyle C_F\frac{C_A}{2}\left(\frac{C_A}{2}-C_F\right)\left(C_A-C_F\right)$&0.5556(2)\\
    \begin{tikzpicture}[baseline=-\the\dimexpr\fontdimen22\textfont2\relax]
      \def\Radius{0.5}
      \draw (0,0) circle (\Radius);
      \draw[color=blue] (30:-\Radius) -- (60:\Radius);
      \draw[color=lellow] (-60:-\Radius) -- (-30:\Radius);
      \draw[color=green] (15:\Radius) -- (-90:\Radius/3);
      \draw[color=cyan] (165:\Radius) -- (-90:\Radius/3);
      \draw[color=red] (-90:\Radius/3) -- (-90:\Radius);
      \useasboundingbox ([shift={(1mm,1mm)}]current bounding box.north east)
            rectangle ([shift={(-1mm,-1mm)}]current bounding box.south west);
    \end{tikzpicture}= $F^e_{cd}{\rm Tr}\left[T^aT^cT^bT^dT^aT^bT^e\right]$
    &$\displaystyle C_F\frac{C_A}{2}\left(\left(\frac{C_A}{2}-C_F\right)
    \left(C_A-C_F\right)-\frac{C_A^2}{2N_c^2}\right)$&-0.4446(2)\\
    \begin{tikzpicture}[baseline=-\the\dimexpr\fontdimen22\textfont2\relax]
      \def\Radius{0.5}
      \def\Rot{-60}
      \draw (0,0) circle (\Radius);
      \draw[color=lellow] (\Rot+15:\Radius) -- (\Rot+135:\Radius);
      \draw[color=blue] (\Rot+210:\Radius) -- (\Rot+90:\Radius);
      \draw[color=cyan] (\Rot+240:\Radius) -- (\Rot-30:\Radius/3);
      \draw[color=green] (\Rot-30:\Radius/3) -- (\Rot+60:\Radius);
      \draw[color=red] (\Rot-30:\Radius/3) -- (\Rot-30:\Radius);
      \useasboundingbox ([shift={(1mm,1mm)}]current bounding box.north east)
            rectangle ([shift={(-1mm,-1mm)}]current bounding box.south west);
    \end{tikzpicture}= $F^e_{bd}{\rm Tr}\left[T^aT^bT^cT^aT^cT^dT^e\right]$
    &$\displaystyle C_F\frac{C_A}{2}\left(\frac{C_A}{2}-C_F\right)^2$&0.0558(3)\\
    \cline{1-3}
    \begin{tikzpicture}[baseline=-\the\dimexpr\fontdimen22\textfont2\relax]
      \def\Radius{0.5}
      \draw (0,0) circle (\Radius);
      \draw[color=blue] (45:-\Radius) -- (45:\Radius);
      \draw[color=red] (315:-\Radius) -- (315:\Radius);
      \useasboundingbox ([shift={(1mm,1mm)}]current bounding box.north east)
            rectangle ([shift={(-1mm,-1mm)}]current bounding box.south west);
    \end{tikzpicture}= ${\rm Tr}\left[T^aT^bT^aT^b\right]$
    &$\displaystyle -C_F\left(\frac{C_A}{2}-C_F\right)$&-0.2221(1)\\
    \begin{tikzpicture}[baseline=-\the\dimexpr\fontdimen22\textfont2\relax]
      \def\Radius{0.5}
      \draw (0,0) circle (\Radius);
      \draw[color=blue] (60:-\Radius) -- (60:\Radius);
      \draw[color=green] (0:-\Radius) -- (0:\Radius);
      \draw[color=red] (300:-\Radius) -- (300:\Radius);
      \useasboundingbox ([shift={(1mm,1mm)}]current bounding box.north east)
            rectangle ([shift={(-1mm,-1mm)}]current bounding box.south west);
    \end{tikzpicture}= ${\rm Tr}\left[T^aT^bT^cT^aT^bT^c\right]$
    &$\displaystyle C_F\left(\frac{C_A}{2}-C_F\right)\left(C_A-C_F\right)$&0.3701(1)\\
    \begin{tikzpicture}[baseline=-\the\dimexpr\fontdimen22\textfont2\relax]
      \def\Radius{0.5}
      \draw (0,0) circle (\Radius);
      \draw[color=blue] (67.5:-\Radius) -- (67.5:\Radius);
      \draw[color=green] (22.5:-\Radius) -- (22.5:\Radius);
      \draw[color=lellow] (-22.5:-\Radius) -- (-22.5:\Radius);
      \draw[color=red] (-67.5:-\Radius) -- (-67.5:\Radius);
      \useasboundingbox ([shift={(1mm,1mm)}]current bounding box.north east)
            rectangle ([shift={(-1mm,-1mm)}]current bounding box.south west);
    \end{tikzpicture}= ${\rm Tr}\left[T^aT^bT^cT^dT^aT^bT^cT^d\right]$
    &$\displaystyle -C_F\left(\left(\frac{C_A}{2}-C_F\right)
    \left(C_A-C_F\right)\left(\frac{3}{2}C_A-C_F\right)-\frac{C_A^3}{4N_c^2}\right)$&-0.1729(1)\\
    \cline{1-3}
  \end{tabular}
  \caption{Selected color coefficients of squared amplitudes in processes with two
    quarks at the leading order, normalized to the common overall factor of $N_c$.
    The colored lines in the diagrams represent gluons, while the black circle
    represents the quark loop. Numerical results have been obtained using
    the algorithm in Sec.~\ref{sec:color_sampling} and are given for $N_c=3$.
    Note that we define $F^a_{bc}=if^{abc}$.
    \label{tab:color_tests}}
\end{table}
The color insertion operators ${\bf T}_i\ldots{\bf T}_j$ in Eq.~\eqref{eq:coll_sub_soft}
are computed using Monte Carlo summation in the color-flow basis.
In the following, we denote the emission of a gluon off parton $i$
by the color branching $(c_i,\bar{c}_i)\to(c_i',\bar{c}_i')(c_g',\bar{c}_g')$,
and the absorption on parton $j$ by the color recombination
$(c_j,\bar{c}_j)(c_g,\bar{c}_g)\to(c_j',\bar{c}_j')$.
We choose to sample the color configuration in the emission
according to the quadratic Casimir operator, ${\bf T}_i^2$.
In the case of gluon emission off a quark, we have
${\bf T}_i^2=T^a_{ik}T^a_{kj}=C_F\delta_{ij}$. In order to allow for a
fully differential sampling of the gluon color index, we insert
a partition of unity in the form $\delta^{ab}=2{\rm Tr}(T^aT^b)$,
leading to $2(T^a_{ik}T^a_{mn})(T^b_{nm}T^b_{kj})$. The first parentheses
are associated with the emission of the gluon, and the second with its absorption.
With the help of the Fierz identity,
$T^a_{ik}T^a_{mn}=1/2(\delta_{in}\delta_{mk}-\delta_{ik}\delta_{mn}/N_c)$,
we identify three distinct color topologies, which are depicted in
Fig.~\ref{fig:color_sampling}. Diagram~(a) squared corresponds to contributions
from $\delta_{in}\delta_{mk}$ only, and carries a weight of $1/2$.
For each given quark color $(c_i,0)$, there are $N_c-1$ such configurations,
leading to a relative weight of $(N_c-1)/(2C_F)$.
Diagram~(b) squared corresponds to contributions from $\delta_{ik}\delta_{mn}$
only, and carries a weight of $1/(2N_c^2)$.
For each given quark color $(c_i,0)$, there are $N_c-1$ such configurations,
leading to a relative weight of $(N_c-1)(1/N_c^2)/(2C_F)$.
Diagrams~(c) squared correspond to contributions from both $\delta_{in}\delta_{mk}$
and $\delta_{ik}\delta_{mn}$, and carry a weight of $(1-1/N_c)^2/2$.
For each given quark color $(c_i,0)$, there is exactly one such configuration,
leading to a relative weight of $(1-1/N_c)^2/(2C_F)$.

Based on these considerations, and a similar decomposition of the quadratic
Casimir operator for gluons, we are eventually led to the following
color sampling algorithm
\begin{enumerate}
\item If the emitter is a quark or an antiquark, assign a weight $C_F$ to the emission
  \begin{enumerate}
  \item With probability $(N_c-1)/(2C_F)$ generate an octet configuration:\\
    if $i$ is a quark / antiquark, choose a new color $c\neq c_i$ \ $c\neq \bar{c}_i$\\
    assign the flow as $(c_i,0)\to(c,0),(c_i,c)$ / $(0,c_i)\to(0,c),(c,c_i)$
  \item With probability $(N_c-1)(1/N_c)^2/(2C_F)$ generate
    a singlet configuration with different colors:\\
    if $i$ is a quark / antiquark, choose a new color $c\neq c_i$  / $c\neq \bar{c}_i$,\\
    set the color indices of the gluon to $(c,c)$
  \item With probability $(1-1/N_c)^2/(2C_F)$ generate
    a singlet configuration with identical colors:\\
    if $i$ is a quark / antiquark, set the color indices of the gluon
    to $(c_i,c_i)$  / $(\bar{c}_i,\bar{c}_i)$
  \end{enumerate}
\item If the emitter is a gluon
  \begin{enumerate}
  \item If $i$ is in an octet state, $c_i\neq\bar{c}_i$, choose a new color $c$
    and assign the emission a weight $N_c$
  \item If $i$ is in a singlet state, $c_i=\bar{c}_i$,
    choose a new color $c\neq c_i$ and assign the emission a weight $N_c-1$
  \item Choose a random permutation,
    either $(c_{i,1},c_{i,2})\to(c_{i,1},c),(c,c_{i,2})$
    or $(c_{i,1},c_{i,2})\to(c,c_{i,1}),(c_{i,2},c)$
  \end{enumerate}
\end{enumerate}
The complete operator ${\bf T}_i\ldots {\bf T}_j$ is restored by
sampling over all possible recombinations of the intermediate
gluon upon insertion of ${\bf T}_j$. The recombination algorithm
proceeds as follows
\begin{enumerate}
\item If the absorber is a quark or antiquark
  \begin{enumerate}
  \item If $j$ is a quark and $c_j=\bar{c}_g$ or
    $\bar{c}_g=c_g$, set the merged color to $(c_g,0)$, else assign weight zero
  \item If $j$ is an antiquark and $\bar{c}_j=c_g$ or
    $c_g=\bar{c}_g$, set the merged color to $(0,\bar{c}_g)$, else assign weight zero
  \end{enumerate}
\item If the absorber is a gluon
  \begin{enumerate}
  \item If $c_g=\bar{c}_j$ and $\bar{c}_g=c_j$, assign weight 2
    and set the merged color randomly to either $(c_j,\bar{c}_g)$ or $(c_g,\bar{c}_j)$
  \item Else if $c_g=\bar{c}_j$ / $\bar{c}_g=c_j$,
    set the merged color to $(c_j,\bar{c}_g)$ / $(c_g,\bar{c}_j)$,
    else assign weight zero
  \end{enumerate}
\end{enumerate}
Note that arbitrarily many insertions may happen before the gluon
emitted by ${\bf T}_i$ is annihilated via ${\bf T}_j$, as required
by Eq.~\eqref{eq:coll_sub_soft}.
The correctness of the above algorithm follows directly from
the decomposition of the generators and the structure constants
of $SU(N_c)$ in the color-flow basis~\cite{Maltoni:2002mq}.
In the context of numerical resummation it is important to note that
the color matrix elements in Eq.~\eqref{eq:single_sudakov} can be
evaluated as a Monte Carlo integral with more than one point per event.
This can be used in practice to improve the convergence of the
overall simulation.

We validate the above algorithm numerically by computing the
color coefficients for gluon webs within a quark loop. They can be
systematically reduced to maximally non-abelian coefficients, which
are related to the quadratic, quartic and higher-point Casimir operators.
This leaves a small number of nontrivial intermediate gluon web configurations,
which need to be evaluated. Table~\ref{tab:color_tests} lists some of these
configurations up to four gluon insertions and compares the analytic results
to Monte Carlo predictions from our algorithm at high statistical accuracy.
In this context we define $F^a_{bc}=if^{abc}$, where $f^{abc}$ are the
SU(3) structure constants.
Note in particular, that the fourth coefficient in the table is related to
the quartic gluon Casimir operator, leading to an additional contribution
of $2/N_c^2$ which arises from double singlet gluon exchange between two gluons.

\section{Numerical results}
\label{sec:numerics}
\begin{figure}[t]
  \centering
  \begin{minipage}{7.5cm}
    \includegraphics[width=\textwidth]{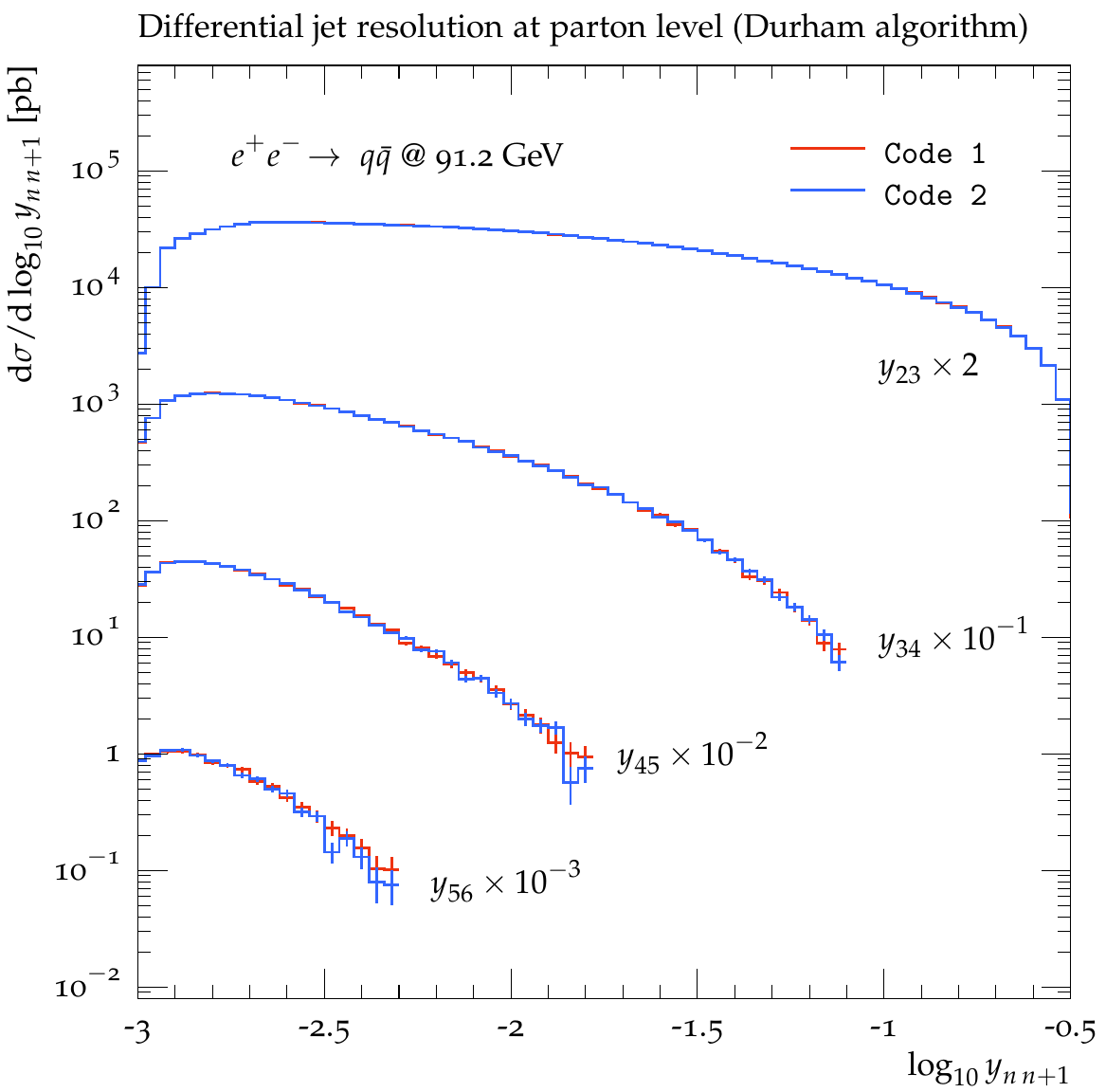}
  \end{minipage}\hskip 5mm
  \begin{minipage}{7.5cm}
    \includegraphics[width=\textwidth]{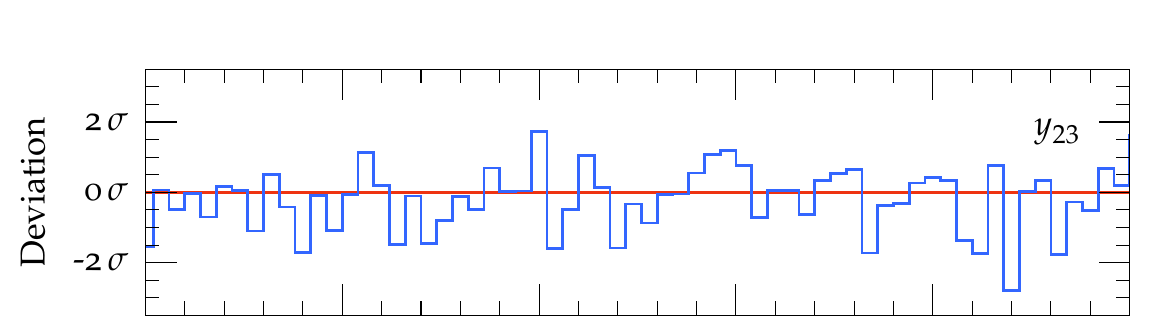}\\[-1mm]
    \includegraphics[width=\textwidth]{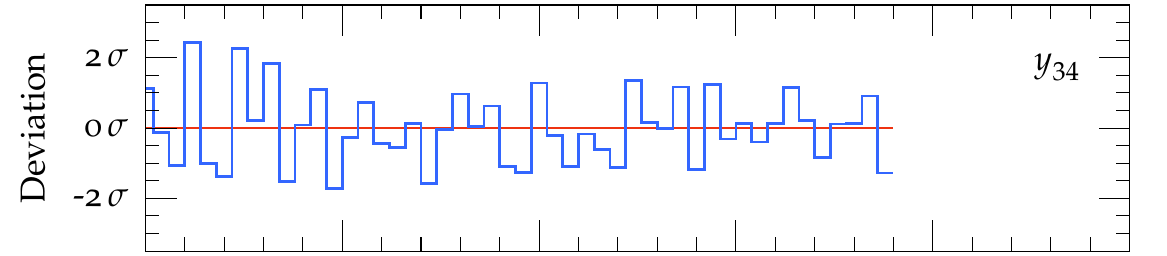}\\[-1mm]
    \includegraphics[width=\textwidth]{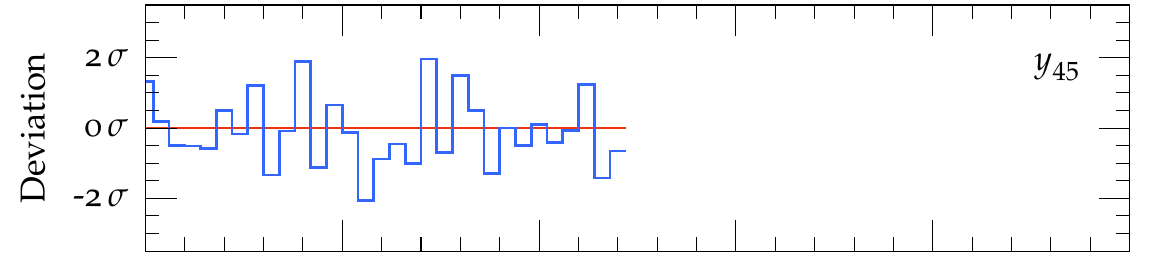}\\[-1mm]
    \includegraphics[width=\textwidth]{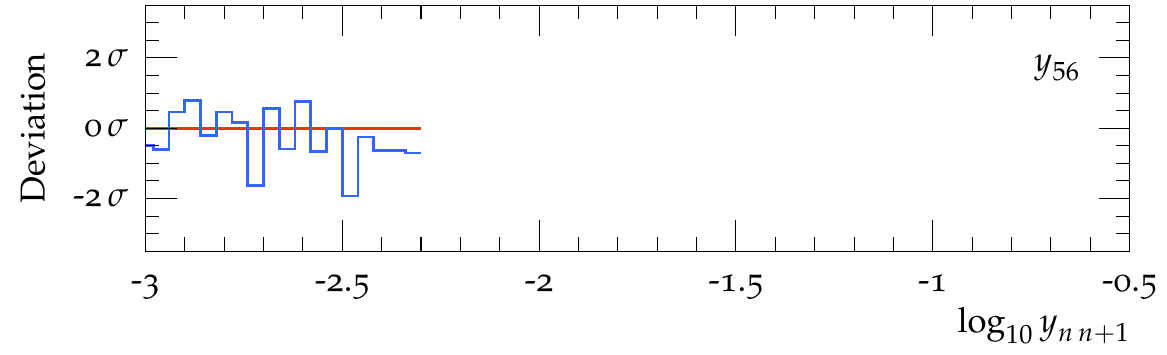}
  \end{minipage}
  \caption{Durham $k_T$-jet resolution scales~\cite{Catani:1991hj} in
    $e^+e^-\to$hadrons at $\sqrt{s}=91.2$~GeV. We compare predictions from two
    independent 
    implementations of our algorithm, labeled ``Code 1'' and ``Code 2''.
    The infrared cutoff is set to $\sqrt{t_c}=\sqrt{t_{c,FC}}=3$~GeV.
    \label{fig:validation}}
\end{figure}
In this section we apply our new algorithm to $e^+e^-\to$~jets at the
$Z$ pole, $\sqrt{s}=91.2~\mathrm{GeV}$. We use a two-loop running coupling
defined by $\alpha_s(Q^2)=0.118$, and the quark mass thresholds
$m_c=1.3$~GeV and $m_b=4.75$~GeV. We cross-check our predictions
using two entirely independent Monte Carlo implementations based
on~\cite{Hoche:2014rga}, which was validated independently
against~\cite{Hoche:2015sya} at high precision. Our simulations
do not include collinear contributions to the splitting functions,
and they are carried out at the parton level. They can therefore
not be compared directly to experimental data. However, they serve
as a first proof-of-concept that color matrix element corrections at arbitrary
multiplicity can be performed in a numerically stable fashion that enables
their application to relevant physics problems in current or past
collider experiments.

We follow the approach in~\cite{Isaacson:2018zdi} and terminate the
subleading color evolution at a scale $t_{c,FC}$ that is insignificantly
larger than the typical parton-shower infrared cutoff of $\sqrt{t_c}\sim 1$~GeV.
All distributions presented here are generated with $\sqrt{t_{c,FC}}=3$~GeV.
We claim that this is not a problem for practical applications, since
hadronization effects typically influence numerical predictions up
to a scale of the order of the $b$-quark mass, and the details of the
fragmentation model have a much larger impact on measurable distributions
in this range than the details of the parton shower. In order to provide
a smooth transition to improved leading-color evolution
below $t_{c,FC}$, we choose a leading color state according to the
probability for a leading-color matrix element to have produced the
partonic final state at scale $t_{c,FC}$. This is similar to how
leading-color configurations are chosen in matching and merging
techniques~\cite{Mangano:2001xp,Hoeche:2009rj}.

Figure~\ref{fig:validation} shows a comparison of predictions for the
Durham $k_T$-jet rates~\cite{Catani:1991hj} in $e^+e^-\to$hadrons at
$\sqrt{s}=91.2$~GeV. Our two independent numerical implementations
of the resummation are statistically compatible and show good convergence,
even in regions of large $k_T$ and for higher jet multiplicity.
\begin{figure}[t]
  \centering
  \includegraphics[width=.45\textwidth]{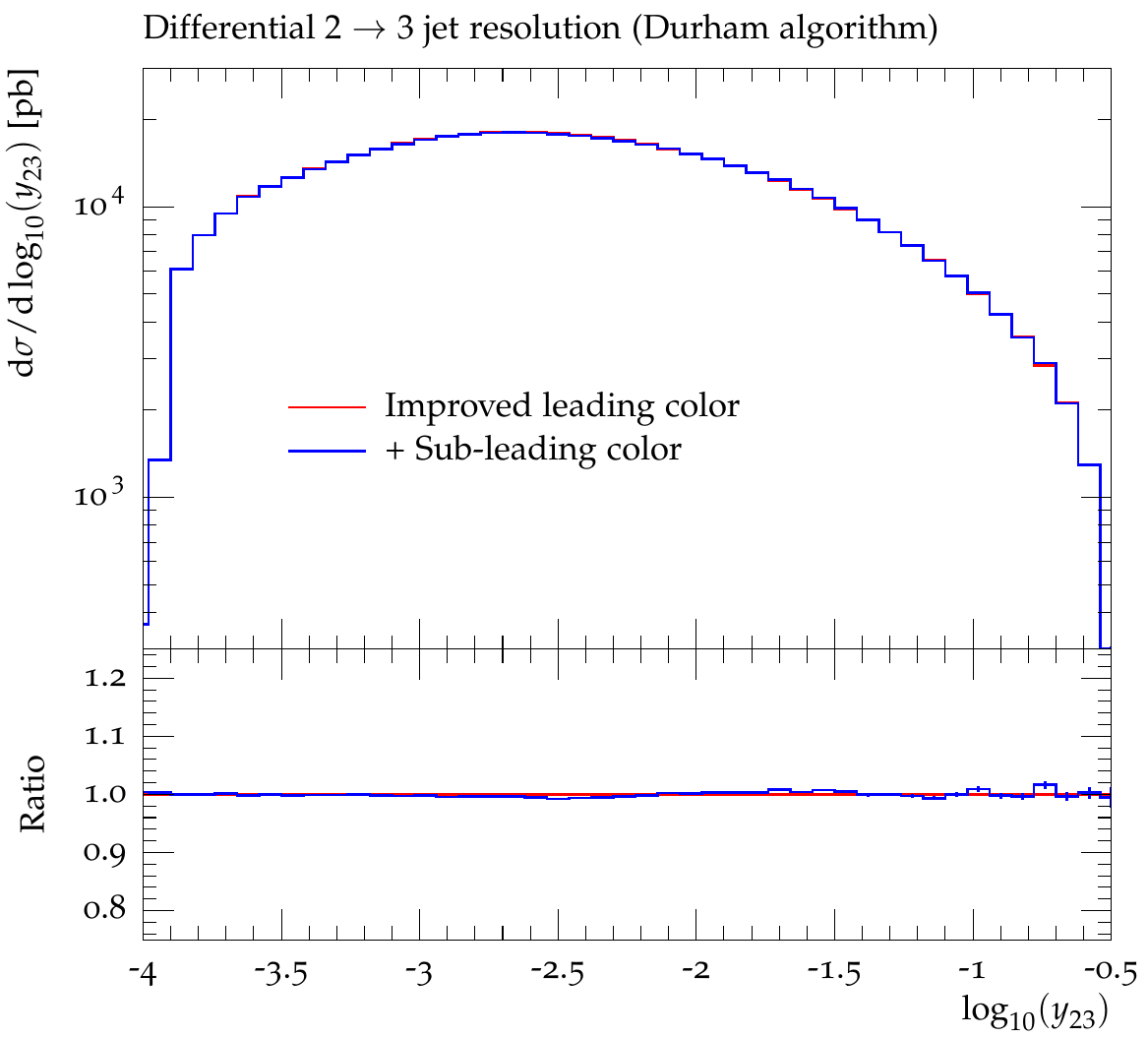}\hskip 3mm
  \includegraphics[width=.45\textwidth]{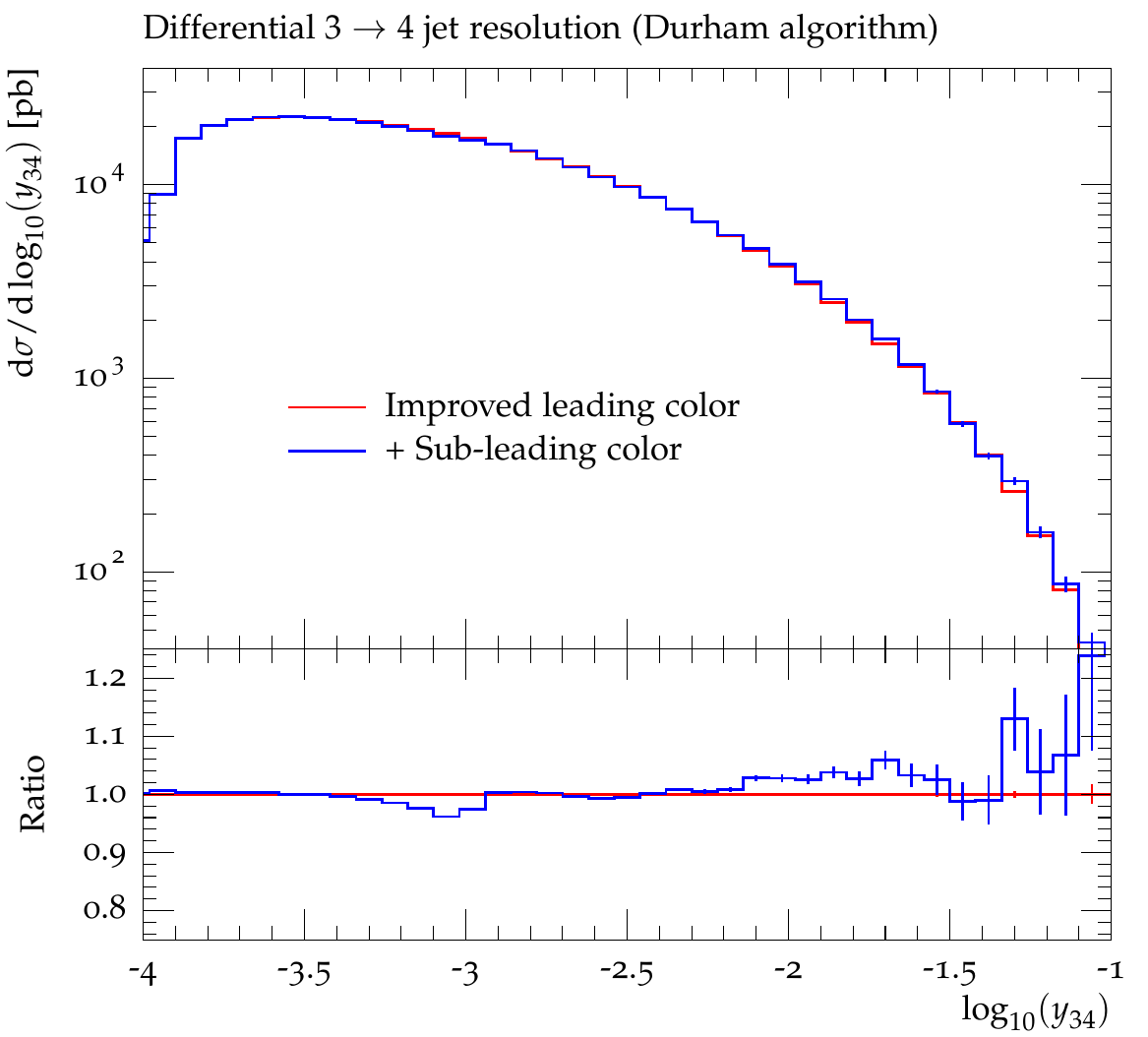}\\
  \includegraphics[width=.45\textwidth]{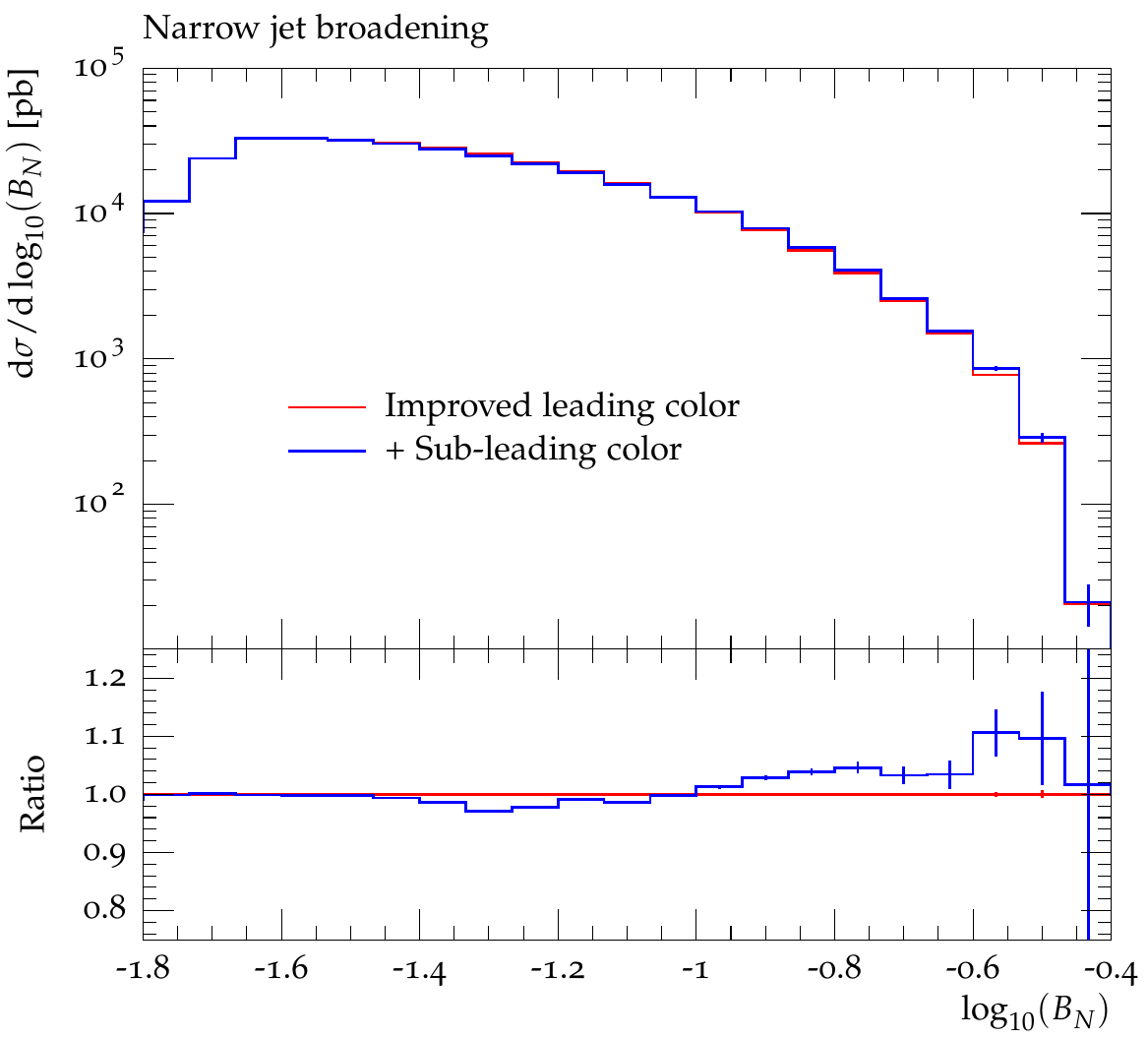}\hskip 3mm
  \includegraphics[width=.45\textwidth]{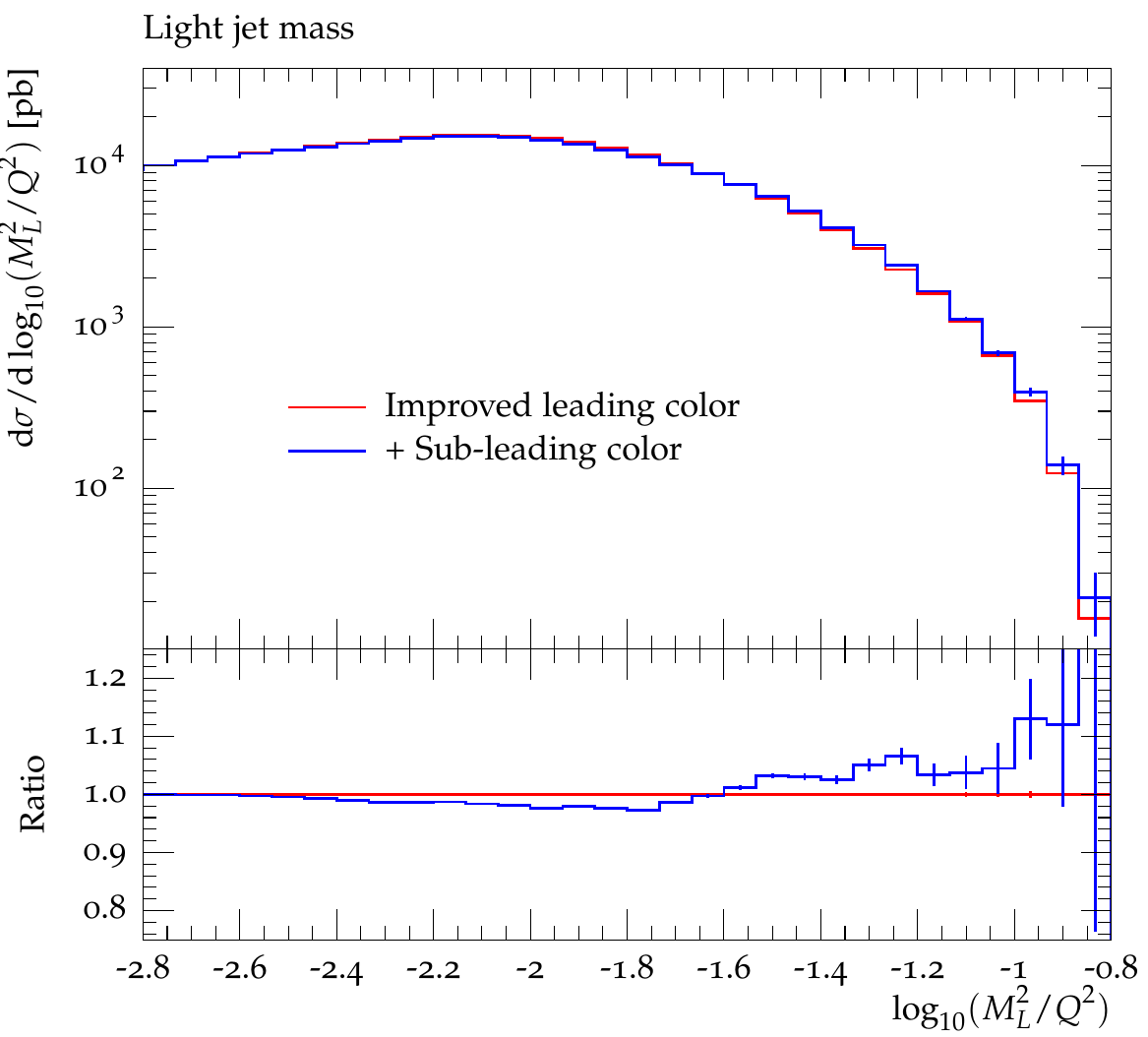}
  \caption{Predictions for Durham $k_T$-jet resolution scales (top), narrow jet
    broadening 
    (bottom left) and light jet mass (bottom right)
    in $e^+e^-\to\text{hadrons}$ at $\sqrt{s}=91.2$~GeV.
    The infrared cutoffs are set to $\sqrt{t_c}=1$~GeV and $\sqrt{t_{c,FC}}=3$~GeV.
    Results using an improved leading color approximation are shown in red,
    and results from the subleading color resummation are in blue.
    \label{fig:shapes}}
\end{figure}
Figure~\ref{fig:shapes} displays numerical predictions at $\sqrt{t_c}=1$~GeV
for the Durham $2\to3$ and $3\to4$ jet scales, and for two nonglobal shape
observables, the narrow jet broadening, $B_N$, and the light jet mass, $M_L$.
We find that the impact of subleading color evolution on all these observables
is less than $10\%$, which agrees with the intuitive notion that
corrections to improved leading color evolution should be of order $1/N_c^2$.
This can be taken as a strong indication that the typically excellent agreement
of modern parton-shower predictions with measured nonglobal shape observables
is not entirely accidental. A variant of the Durham $n\to(n+1)$ jet scales has
been resummed recently and matched to next-to-leading order (NLO), to achieve
$\mathrm{NLO}+\mathrm{NLL^\prime}$ accuracy in \cite{Baberuxki:2019ifp},
including a quantification of prospective subleading color contributions. Our
results are compatible with the smallness of the effects observed there, in
particular when noting that the results in \cite{Baberuxki:2019ifp} are matched
to a fixed order NLO calculation relative to the Born process while we present
pure parton shower results here.

\section{Conclusions}
\label{sec:conclusions}
We have presented a novel Monte Carlo method for soft-gluon resummation that
allows one to generate parton-level events and can be incorporated into existing
parton showers in order to improve their formal precision.
Along with this manuscript, we provide a proof-of-concept implementation
that can be used for numerical studies in $e^+e^-\to$hadrons.
We find that the impact of subleading color evolution on the Durham $k_T$-jet scales,
on narrow jet broadening, and on the light jet mass agrees with the naive
expectation that corrections to existing parton-shower approaches should be
suppressed by $\mathcal{O}(1/N_c^2)$.
It will be interesting to investigate the impact on dedicated observables,
which probe nontrivial color correlations, as for example in~\cite{Larkoski:2019fsm}.
In this work we have neglected the matrix structure of soft virtual corrections.
A future study will address the feasibility of including these contributions as well.

\section*{Acknowledgments}
We thank Joshua Isaacson for numerous fruitful discussions on the stochastic sampling
of color configurations.
We are grateful to Simone Marzani for his comments on the manuscript,
and for many stimulating discussions on soft-gluon resummation. DR
thanks Steffen Schumann for his support.

We are particularly grateful to Dave Soper and Zoltan Nagy, as well as Jeff Forshaw, Simon Pl{\"a}tzer
and Jack Holguin for pointing out that our original claim to achieve resummation at full color accuracy
does not hold due to the unitarity argument in Eq.~\eqref{eq:full_sudakov}~\cite{Holguin:2020xx}.

This work used resources of the Fermi National Accelerator Laboratory (Fermilab),
a U.S. Department of Energy, Office of Science, HEP User Facility.
Fermilab is managed by Fermi Research Alliance, LLC (FRA), acting under
Contract No. DE--AC02--07CH11359.
DR acknowledges support from the German-American Fulbright Commission
which funded his research visit to Fermilab. DR further acknowledges funding
from the European Union’s Horizon 2020 research and innovation program as part
of the Marie Skłodowska-Curie Innovative Training  Network MCnetITN3 (grant
agreement no. 722104) and from BMBF (contract 05H18MGCA1). 

\appendix
\section{Relation to analytic resummation of global event shapes}
\label{sec:caesar}
A general framework for resumming a large class of observables has been
developed in the context of the program Caesar~\cite{Banfi:2001bz,Banfi:2003je,
  Banfi:2004nk,Banfi:2004yd}. This formalism, in particular including the
computation of color structures for in principle arbitrary multiplicities, was
automated in~\cite{Gerwick:2014gya} and recently applied to resummed
calculations in $e^+e^-\to\mathrm{jets}$ \cite{Marzani:2019evv,Baberuxki:2019ifp}.
In this formalism, color coherence is an integral part in treating multiple
real emissions. We therefore find it relevant to test how this is reproduced
in our algorithm.

For two emissions at large rapidity difference, color coherence implies that
${\bf\Gamma}_{n+1}$ should reduce to ${\bf\Gamma}_n$. We are thus left to compute
the following difference
\begin{equation}
  \Delta{\bf\Gamma}_{n+1}({\bf\Gamma})=
    {\bf\Gamma}_n({\bf\Gamma}_{n+1}({\bf\Gamma}))
    -{\bf\Gamma}_n({\bf\Gamma}_n({\bf\Gamma}))
    =-\sum_{i,k=1}^n\sum_{\substack{j=1\\j\neq i}}^n
    {\bf T}_i\Big({\bf T}_{n+1}{\bf\Gamma}\,{\bf T}_k+
    {\bf T}_k{\bf\Gamma}\,{\bf T}_{n+1}\Big)\,{\bf T}_j\,w_{n+1k}\,w_{ij}\;.
\end{equation}
We rewrite this as
\begin{equation}
  \Delta{\bf\Gamma}_{n+1}({\bf\Gamma})
  =-\sum_{i=1}^n\sum_{\substack{j=1\\j\neq i}}^n
  \sum_{k=1}^n\sum_{\substack{l=1\\l\neq k}}^n
      {\bf T}_i{\bf T}_k\Delta_n^{(kl)}({\bf\Gamma})\,
      {\bf T}_l{\bf T}_j\,w_{kl}\,w_{ij}\;,
\end{equation}
where
\begin{equation}
  \Delta_n^{(kl)}({\bf\Gamma})=
    \frac{{\bf T}_{n+1}{\bf\Gamma}\,{\bf T}_l}{
      {\bf T}_k{\bf\Gamma}\,{\bf T}_l}\,X_{n+1q}^{kl}+
    \frac{{\bf T}_k{\bf\Gamma}\,{\bf T}_{n+1}}{
      {\bf T}_k{\bf\Gamma}\,{\bf T}_l}\,X_{n+1q}^{kl}\;,
\end{equation}
and where we have defined the cross ratio
\begin{equation}\label{eq:cross_ratio}
  X_{ij}^{kl}=\frac{w_{ik}}{w_{kl}}=
  \frac{s_{ik}s_{lj}}{s_{ij}s_{kl}}\;.
\end{equation}
Note that $X_{ij}^{kl}$ does not scale if any of the particles becomes soft.
We can parametrize it in terms of the rapidity difference
$\Delta\eta_{kl}^{(ij)}$ and azimuthal angle difference $\Delta\phi_{kl}^{(ij)}$
with respect to the light-cone directions defined by $p_i$ and $p_j$, 
\begin{equation}
  \cosh\Delta\eta_{kl}^{(ij)}=\frac{s_{ik}s_{jl}+s_{il}s_{jk}}{
    \sqrt{s_{ik}s_{jk}\,s_{il}s_{jl}}}\;,\qquad
  \cos\Delta\phi_{kl}^{(ij)}=\frac{s_{ik}s_{jl}+s_{il}s_{jk}-s_{ij}s_{kl}}{
    \sqrt{s_{ik}s_{jk}\,s_{il}s_{jl}}}\;,
\end{equation}
such that
\begin{equation}
  \begin{split}
    X_{ij}^{kl}=\frac{\cosh\Delta\eta_{kl}^{(ij)}-\sinh\Delta\eta_{kl}^{(ij)}}{
      \cosh\Delta\eta_{kl}^{(ij)}-\cos\Delta\phi_{kl}^{(ij)}}\;.
  \end{split}
\end{equation}
Averaging over the azimuthal angle gives
\begin{equation}
  \begin{split}
    \bar{X}_{ij}^{kl}
    =\frac{1}{2\pi}\int_{0}^{2\pi}{\rm d}\Delta\phi_{kl}^{(ij)}\,X_{ij}^{kl}
    =\coth\Delta\eta_{kl}^{(ij)}-1\;.
  \end{split}
\end{equation}
For the global, recursively infrared and collinear safe observables
considered in~\cite{Banfi:2001bz,Banfi:2003je,Banfi:2004nk,Banfi:2004yd},
the region $\Delta\eta_{iq}\ll 1$ is completely described by implementing
the collinear evolution of gluon webs to the desired accuracy, which
only depends on (the quadratic color Casimir operator of) the original hard leg.
The remaining groups of gluons have $1/\Delta\eta\approx \alpha_s\ln(1/\epsilon)\ll 1$.
In this case, $\bar{X}_{ij}^{kl}$ vanishes, and we obtain
$\int{\rm d}\phi\,\Delta_{n+1}({\bf\Gamma})\to 0$.
We hence correctly reproduce the picture of~\cite{Banfi:2004yd}: The
radiation of an additional soft gluon can either be computed in the collinear
limit, or it can be described using the original soft anomalous dimension,
ignoring the change in color flow arising from previous soft-gluon insertions.
A similar description of this effect is obtained in the coherent branching
formalism~\cite{Ellis:1991qj}.

\section{Explicit examples of insertion operators}
\label{sec:examples}
In this appendix we demonstrate the application of the soft-gluon insertion
formula, Eq.~\eqref{eq:coll_sub_soft} using two simple examples.
The two-parton case being trivial, we investigate soft insertions
into three- and four-parton matrix elements, as they occur, for example,
in $e^+e^-\to$ hadrons or $h\to gg$ decays.
Due to crossing invariance of the hard matrix elements, these examples also
cover the highly relevant cases of charged and neutral current Drell-Yan
and Higgs-boson production at hadron colliders, as well as charged
and neutral current deep inelastic scattering.

\subsection{Three radiators}
All elements of the color algebra can be expressed in terms of the quadratic
Casimir operators ${\bf T}_1^2$, ${\bf T}_2^2$ and ${\bf T}_3^2$ by means of
color charge conservation, ${\bf T}_1+{\bf T}_2+{\bf T}_3=0$.
The remaining insertion operators can be written as
\begin{equation}\label{eq:colors_three_partons}
  \begin{split}
    {\bf T}_1{\bf T}_2=\;&\frac{1}{2}(-C_1-C_2+C_3)\;,\\
    {\bf T}_1{\bf T}_3=\;&\frac{1}{2}(-C_1+C_2-C_3)\;,\\
    {\bf T}_2{\bf T}_3=\;&\frac{1}{2}(C_1-C_2-C_3)\;.
  \end{split}
\end{equation}
Based on Eq.~\eqref{eq:coll_sub_soft}, the complete soft insertion operator
with 1,2 being the same type of parton (either quark or gluon), is then given by
\begin{equation}
  \begin{split}
    {\bf\Gamma}_3({\bf 1})=&\;\frac{1}{2}\sum_{i=1}^3\sum_{\substack{j=1\\j\neq i}}^3\bigg(
    {\bf T}^2_iP^i_j
    +\sum_{\substack{k=j+1\\k\neq i,j}}^3{\bf T}_i{\bf T}_j\tilde{P}^i_{kj}\bigg)\\
    =&\;\frac{1}{2}\bigg(C_1\big(P^1_2+P^1_3+P^2_1+P^2_3\big)+C_3\big(P^3_1+P^3_2\big)
    +\bigg(\frac{C_3}{2}-C_1\bigg)\big(\tilde{P}^1_{32}+\tilde{P}^2_{31}\big)
    -\frac{C_3}{2}\big(\tilde{P}^1_{23}+\tilde{P}^2_{13}\big)\bigg)\\
    =&\;C_1w_{12}+\frac{C_3}{2}\Big(w_{13}+w_{23}-w_{12}\Big)\;.
  \end{split}
\end{equation}

\subsection{Four radiators}
We choose ${\bf T}_1^2$, \ldots, ${\bf T}_4^2$, ${\bf T}_1{\bf T}_4$
and ${\bf T}_1{\bf T}_3$ to be the independent elements of the color algebra.
The remaining insertion operators can be expressed in terms of these operators
as
\begin{equation}\label{eq:colors_four_partons}
  \begin{split}
    {\bf T}_1{\bf T}_2=&\;-C_1-{\bf T}_1{\bf T}_3-{\bf T}_1{\bf T}_4\;,\\
    {\bf T}_2{\bf T}_3=&\;\frac{1}{2}\Big(C_1-C_2-C_3+C_4\Big)+{\bf T}_1{\bf T}_4\;,\\
    {\bf T}_2{\bf T}_4=&\;\frac{1}{2}\Big(C_1-C_2+C_3-C_4\Big)+{\bf T}_1{\bf T}_3\;,\\
    {\bf T}_3{\bf T}_4=&\;\frac{1}{2}\Big(-C_1+C_2-C_3-C_4\Big)-{\bf T}_1{\bf T}_3-{\bf T}_1{\bf T}_4\;.
  \end{split}
\end{equation}
Based on Eq.~\eqref{eq:coll_sub_soft}, the complete soft insertion operator
for the four parton case with 1,2 and 3,4 being the same type of parton
(either quark or gluon), then reads
\begin{equation}
  \begin{split}
    {\bf\Gamma}_4({\bf 1})=&\;\frac{1}{3}\sum_{i=1}^{4}\sum_{\substack{j=1\\j\neq i}}^4
    \Bigg({\bf T}^2_iP^i_j
    +\sum_{\substack{k=j+1\\k\neq i,j}}^4{\bf T}_i{\bf T}_j\tilde{P}^i_{kj}\Bigg)\\
    =&\;\frac{1}{3}\bigg(C_1\big(P^1_2+P^1_3+P^1_4+P^2_1+P^2_3+P^2_4\big)
    +C_3\big(P^3_1+P^3_2+P^3_4+P^4_1+P^4_2+P^4_3\big)\\
    &\qquad+{\bf T}_1{\bf T}_2\big(\tilde{P}^1_{32}+\tilde{P}^1_{42}+\tilde{P}^2_{31}+\tilde{P}^2_{41}\big)
    +{\bf T}_1{\bf T}_3\big(\tilde{P}^1_{23}+\tilde{P}^1_{43}+\tilde{P}^3_{21}+\tilde{P}^3_{41}\big)\\
    &\qquad+{\bf T}_1{\bf T}_4\big(\tilde{P}^1_{24}+\tilde{P}^1_{34}+\tilde{P}^4_{21}+\tilde{P}^4_{31}\big)
    +{\bf T}_2{\bf T}_3\big(\tilde{P}^2_{13}+\tilde{P}^2_{43}+\tilde{P}^3_{12}+\tilde{P}^3_{42}\big)\\
    &\qquad+{\bf T}_2{\bf T}_4\big(\tilde{P}^2_{14}+\tilde{P}^2_{34}+\tilde{P}^4_{12}+\tilde{P}^4_{32}\big)
    +{\bf T}_3{\bf T}_4\big(\tilde{P}^3_{14}+\tilde{P}^3_{24}+\tilde{P}^4_{13}+\tilde{P}^4_{23}\big)\bigg)\\
    =&\;\frac{1}{3}\bigg(C_1\big(P^1_2+P^1_3+P^1_4+P^2_1+P^2_3+P^2_4\big)
    +C_3\big(P^3_1+P^3_2+P^3_4+P^4_1+P^4_2+P^4_3\big)\\
    &\qquad+3{\bf T}_1{\bf T}_3\big(\tilde{P}^1_{23}+\tilde{P}^2_{14}+\tilde{P}^3_{41}+\tilde{P}^4_{32}\big)
    +3{\bf T}_1{\bf T}_4\big(\tilde{P}^1_{24}+\tilde{P}^2_{13}+\tilde{P}^3_{42}+\tilde{P}^4_{31}\big)\\
    &\qquad-C_1\big(\tilde{P}^1_{32}+\tilde{P}^1_{42}+\tilde{P}^2_{31}+\tilde{P}^2_{41}\big)
    -C_3\big(\tilde{P}^3_{14}+\tilde{P}^4_{23}+\tilde{P}^4_{13}+\tilde{P}^4_{23}\big)\bigg)\\
    =&\;C_1w_{12}+C_3w_{34}
    +{\bf T}_1{\bf T}_3\big(w_{12}+w_{34}-w_{13}-w_{24}\big)
    +{\bf T}_1{\bf T}_4\big(w_{12}+w_{34}-w_{14}-w_{23}\big)\;.
  \end{split}
\end{equation}

\bibliography{journal}
\end{document}